\documentclass[10pt,conference,a4paper]{IEEEtran}

\usepackage[utf8]{inputenc}
\usepackage{cite}
\usepackage{amsmath,amssymb}
\usepackage{graphicx}
\usepackage{capt-of}
\usepackage{booktabs}
\usepackage{multirow}
\usepackage{xcolor}
\usepackage{colortbl}
\usepackage{url}
\usepackage[hidelinks]{hyperref}
\usepackage{xspace}
\usepackage{balance}

\newcommand{\tool}{Bug2Scenario\xspace}
\newcommand{\gain}[2]{\textbf{#1}{\scriptsize\textcolor{green!50!black}{$_{+#2}$}}}
\newcommand{\drop}[2]{#1{\scriptsize\textcolor{orange!80!black}{$_{#2}$}}}
\newcommand{\same}[1]{#1{\scriptsize\textcolor{gray}{$_{\phantom{+0}0.0}$}}}

\title{\tool: Root-Cause-Preserving Test Scenario Amplification for Autonomous Driving Software}

\author{
\IEEEauthorblockN{Yuchen Cao\textsuperscript{1,\textdagger}, Zhenyu Mao\textsuperscript{1,\textdagger}, Jacky Wai Keung\textsuperscript{1},
Yifei Wang\textsuperscript{1}, Yihan Liao\textsuperscript{1}, and Jingyu Zhang\textsuperscript{2,*}}
\IEEEauthorblockA{\textsuperscript{1}City University of Hong Kong, Hong Kong\\
\{cynthcao2-c, zhenyumao2-c, ywang4748-c, yihanliao4-c\}@my.cityu.edu.hk, Jacky.Keung@cityu.edu.hk}
\IEEEauthorblockA{\textsuperscript{2}Hong Kong Metropolitan University, Hong Kong\\
\textsuperscript{\textdagger}Equal contribution. \textsuperscript{*}Corresponding author: fzhang@hkmu.edu.hk}
}

\begin{document}
\bstctlcite{IEEEexample:BSTcontrol}

\maketitle

\begin{abstract}
Simulation-based fuzzing exposes autonomous driving system (ADS) failures by repeatedly mutating and executing scenarios, but converting one failure into reusable regression scenarios remains difficult. Such variants test whether the mechanism persists after software changes, for example by varying weather or actor placement while retaining a collision's actor--route conflict. We present \tool, a root-cause-guided post-fuzzing framework. From scenario data, traces, and oracle evidence, it extracts a root-cause pattern. An LLM proposes semantic variations over actors, missions, routes, and weather. Deterministic rules construct DriveFuzz-compatible scenarios. Execution evidence and manual review assess preservation. Separating these responsibilities prevents fluency, syntax, or recurring labels from masquerading as causal evidence. Under matched seeds, six LLM backbones, candidate budget, and execution pipeline, \tool improves strict preservation by 6.7 percentage points ($p=0.104$; not significant) and boundary-inclusive preservation by 10.6 points ($p=0.011$). Feedback helps some initially unsuccessful seeds yield a strictly preserved candidate. Seed-level stopping then avoids further API calls and simulator runs. Remaining failures reflect actor-placement constraints, simulator instability, and incomplete or conflicting causal evidence. \tool thus produces reliable ADS regression assets. Our implementation is available at{\footnotesize\href{https://github.com/YgcCoder/Bug2Sce}{\mbox{\texttt{https://github.com/YgcCoder/Bug2Sce}}}}.
\end{abstract}


\section{Introduction}
Fuzz testing in simulation is a central technique for exercising autonomous driving systems (ADSs) because safety-critical interactions are expensive, rare, and difficult to reproduce on public roads~\cite{liao2025advancing,lou2022testing}. CARLA and Autoware provide controllable cyber-physical test environments in which engineers can vary roads, actors, missions, and weather while observing a production-scale driving stack~\cite{dosovitskiy2017carla,kato2018autoware}. Automated methods built on these platforms search large scenario spaces for collisions, traffic-law violations, perception errors, and unsafe vehicle states.

Failure discovery is only the beginning of testing. After a fuzzer exposes a violation, engineers need regression assets that reproduce it under controlled variation. Repeated execution does not reveal which scene details are incidental or required. A useful test family must support debugging, regression testing, and maintenance while keeping the failure boundary interpretable.

Most ADS testing research emphasizes discovery. DeepTest and DeepRoad transform driving scenes, whereas AV-Fuzzer and DriveFuzz mutate executable scenarios~\cite{tian2018deeptest,zhang2018deeproad,li2020av,kim2022drivefuzz}. Recent methods further improve search efficiency, diversity, or responsibility-aware violation discovery~\cite{haq2022efficient,lu2022learning,cheng2023behavexplor,huai2023doppelganger}. Post-fuzzing test amplification instead starts from a failure-triggering seed and generates controlled variants to preserve its mechanism rather than discover an arbitrary new failure. Its output is not a new fault report, but an auditable set of regression candidates traceable to the original evidence. It asks \emph{which variants reproduce the mechanism of a failure already found?}

\begin{figure}[!t]
    \centering
    \includegraphics[width=\columnwidth]{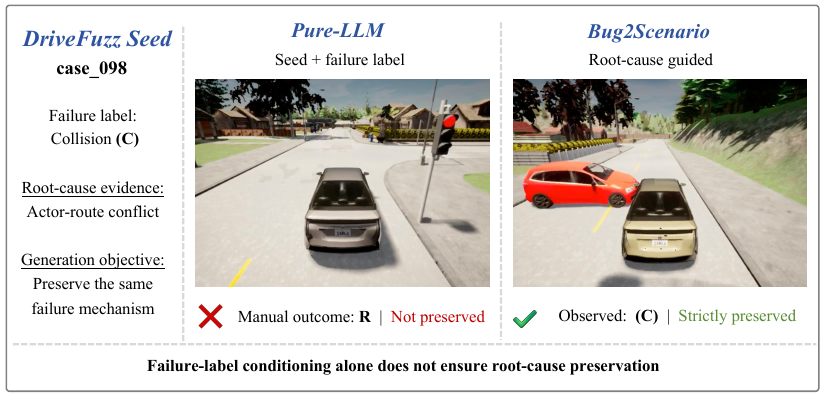}
    \caption{Motivating example from \texttt{case\_098}. The seed failure is a \textbf{collision} caused by an actor--route conflict. Failure-label conditioning alone produces a \textbf{red-light violation} and does not preserve the mechanism, whereas root-cause-guided generation retains the \textbf{collision} mechanism. Thus, matching or triggering an oracle outcome is not sufficient evidence of root-cause preservation.}
    \label{fig:motivating-example}
\end{figure}

The central engineering gap is the distinction between symptom recurrence and mechanism preservation. A failure oracle is a detection rule. Its outcome is the \textbf{collision (C)}, \textbf{red-light violation (R)}, \textbf{lane invasion (L)}, or \textbf{stuck/immobility (S)} label from one run. Symptom recurrence repeats a compatible outcome, whereas mechanism preservation also retains the seed's critical interaction and failure progression. In Fig.~\ref{fig:motivating-example}, \texttt{case\_098} is an actor--route collision. Label-only conditioning instead yields a red-light violation, while root-cause guidance retains that collision mechanism. The same outcome can still arise from different interactions. Parseability, executability, and a matching outcome respectively show that JSON can be read, a scenario can run, and a visible symptom recurs. None alone proves mechanism preservation.

Post-fuzzing test amplification consequently needs explicit preservation constraints and graded assessment. Unconstrained variation may add an unrelated failure to the same bug family, while a scenario near a behavioral boundary may retain the critical interaction but weaken its visible symptom. Binary acceptance based only on syntax or matching outcomes cannot distinguish these outcomes or support reliable scenario-level regression asset generation. Grading also records partial evidence without admitting it as strict reproduction.

LLMs propose semantic variation but create distinct schema, execution, and preservation risks. They support search, fuzz-input generation, and execution-guided repair~\cite{lemieux2023codamosa,deng2023large,xia2024fuzz4all,chen2024chatunitest}, while language and multimodal models propose driving scenarios~\cite{zhang2024chatscene,yang2023llm4drive,cui2024survey}. In LLM-assisted fuzz-input generation, a model proposes tests or structured configurations for another stage to execute and validate. A fluent proposal may place an actor outside a drivable region, omit a critical mission field, or depart from the seed mechanism. Existing fuzzers target discovery, structured generation does not test membership in a seed's bug family, and direct LLM generation guarantees neither schema compliance nor mechanism preservation. One response should not control proposal, coordinates, and causal judgment. Separating these roles makes rejection attributable to generation, construction, execution, or preservation assessment.

In this work, we present \tool as an LLM-assisted but validator-centered approach to post-fuzzing test amplification. Seed evidence defines a root-cause pattern, and an LLM proposes high-level variations over semantic conditions such as actors, missions, routes, and weather rather than final simulator coordinates or JSON. Deterministic concretization then produces DriveFuzz-compatible JSON. Schema checks and execution artifacts support manual assessment of invalid, no-failure, different-failure, boundary, and strictly preserved outcomes. Structural errors return to concretization, whereas an observed execution that conflicts with the seed mechanism revises the causal explanation or later generation constraints. Validated patterns remain anchored to the original seed rather than accumulated feedback. Under matched generation and execution budgets, \tool improves strict and boundary-inclusive preservation over Pure-LLM. Same-seed feedback repairs some first-pass failures before recovery plateaus. The resulting test families help engineers study failure boundaries without mixing unrelated mechanisms. Our contributions are:
\begin{itemize}
    \item We introduce a root-cause-preserving test-generation framework for post-fuzzing amplification, distinguishing same-mechanism reproduction from failure discovery and visible violation recurrence.

    \item \tool provides an evidence-grounded workflow that combines root-cause pattern extraction, LLM-assisted high-level variation, deterministic concretization, and execution-based mechanism preservation assessment.

    \item We present a matched empirical evaluation across six LLM backbones and a Pure-LLM baseline, reporting strict and boundary-inclusive preservation separately.

    \item We analyze iterative same-seed feedback and cost control, identifying where execution and manual-review evidence repairs seeds, where recovery plateaus, and how seed-level stopping avoids later calls and simulations.
\end{itemize}

\begin{figure*}[!t]
    \centering
    \includegraphics[width=0.98\textwidth]{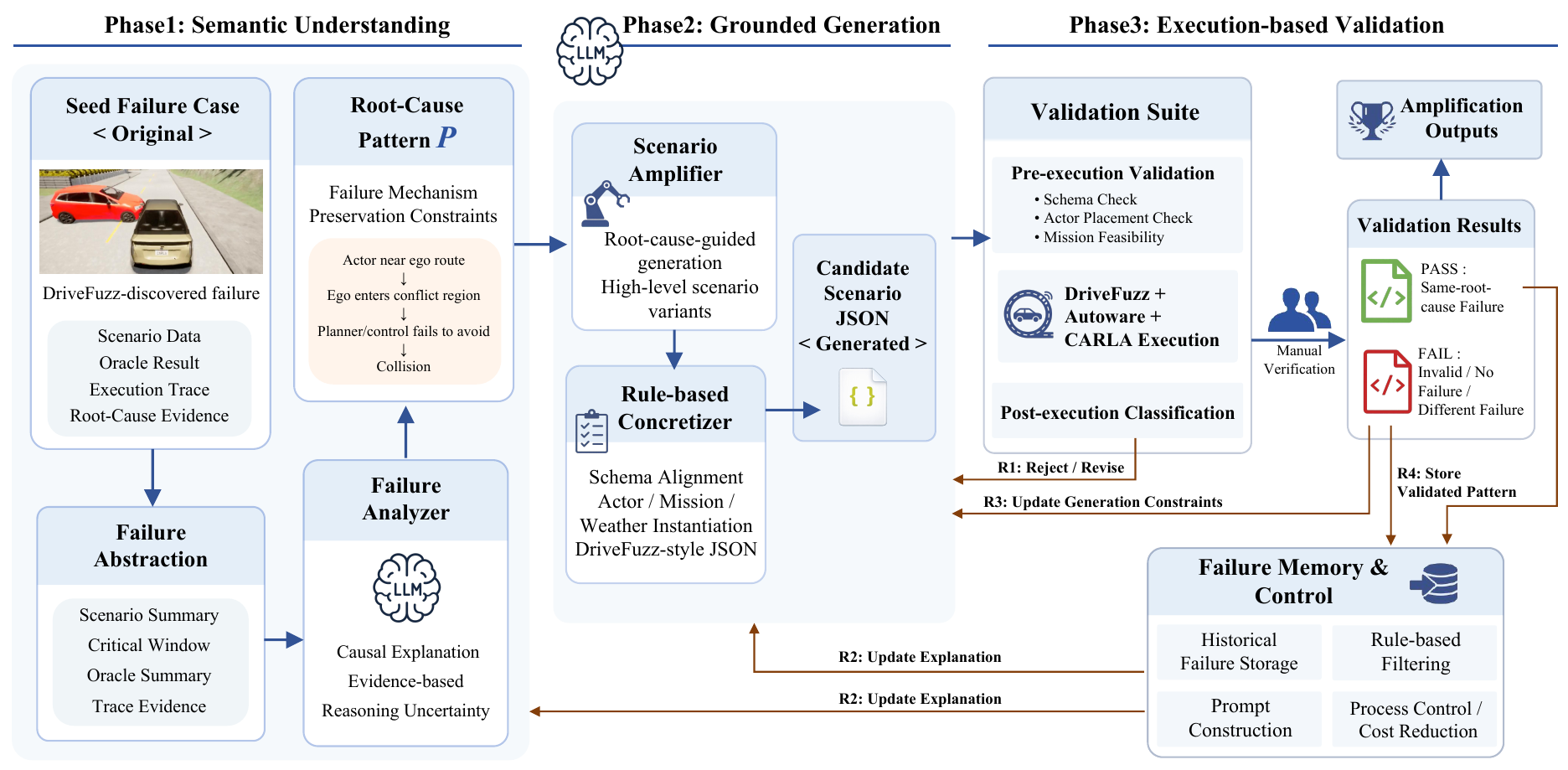}
    \caption{Overview of \tool. The workflow consists of semantic understanding, grounded generation, and execution-based validation. Starting from a DriveFuzz-discovered seed failure, \tool abstracts scenario evidence, uses LLM-assisted causal explanation to extract a root-cause pattern, generates guided candidate scenarios, concretizes them into DriveFuzz-style JSON, and validates them through pre-execution checks and DriveFuzz, Autoware, and CARLA execution. Feedback updates explanations, generation constraints, and failure memory while retaining the seed pattern as the target.}
    \label{fig:bug2scenario-overview}
\end{figure*}

\section{Background and Related Work}

\subsection{Software Testing and Scenario Generation in ADS}

ADS testing spans learned-component checks and system-level scenario execution. Component methods include DeepTest~\cite{tian2018deeptest}, DeepRoad~\cite{zhang2018deeproad}, DeepHunter~\cite{xie2019deephunter}, DeepXplore~\cite{pei2017deepxplore}, DLFuzz~\cite{guo2018dlfuzz}, and Surprise Adequacy~\cite{kim2019guiding}. Recent work evaluates test-selection metrics under distribution shifts~\cite{zhang2026empirical} and probes driving models using transferable~\cite{zhang2024enhancing} or adaptive multiobjective attacks~\cite{zhang2024uniada}. System-level search includes AV-Fuzzer~\cite{li2020av}, DriveFuzz~\cite{kim2022drivefuzz}, SAMOTA~\cite{haq2022efficient}, DeepCollision~\cite{lu2022learning}, BehAVExplor~\cite{cheng2023behavexplor}, and DoppelTest~\cite{huai2023doppelganger}. Unlike component tests, these executions must satisfy map, mission, traffic, and simulator constraints. Scenic~\cite{fremont2019scenic}, VerifAI~\cite{dreossi2019verifai}, LawBreaker~\cite{sun2022lawbreaker}, and AsFault~\cite{gambi2019asfault} support specification, falsification, or scenario generation. They target property violations or failure discovery, not preservation of a selected seed mechanism. Coverage, diversity, and oracle yield therefore cannot define regression-family membership. Regression scenarios need a stable intent linking execution to seed evidence; \tool retains DriveFuzz's representation and checks candidates against that mechanism.

Traditional test amplification starts from existing tests and transforms their inputs or assertions to expose additional behavior; DSpot, for example, improves developer-written Java unit tests through input and assertion amplification~\cite{danglot2019dspot}. Recent LLM-assisted ADS testing instead uses semantic knowledge to guide scenario discovery: SLaFE translates traffic laws into structured constraints and combines them with cooperative evolutionary fuzzing to search for violation-triggering scenarios~\cite{liu2025slafe}. These approaches respectively amplify program-level tests or discover law-violating driving scenarios. In contrast, \tool starts from one executed ADS failure and constructs a regression family whose candidates must retain that seed's interaction mechanism.

\subsection{Failure Oracles and Mechanism Preservation}

DriveFuzz~\cite{kim2022drivefuzz}, AV-Fuzzer~\cite{li2020av}, and LawBreaker~\cite{sun2022lawbreaker} use oracles to identify violations, not execution equivalence. Two collisions may arise from different actor interactions, and one interaction may approach a boundary where its symptom weakens. Regression membership instead depends on the seed's actor--route geometry, mission segment, temporal behavior, system response, and critical window. Graded outcomes record uncertainty when incomplete traces retain only part of this relation. \tool combines a seed pattern with manual assessment to distinguish strict, boundary, and non-preserved outcomes without treating recurrence as proof.

\subsection{LLM-assisted Software Testing}

LLM-assisted software engineering raises data-quality and output-validation concerns. At repository level, HGARena organizes budgeted multi-agent issue resolution around executable tests as the final acceptance signal~\cite{cao2026hgarena}. Although HGARena targets patch resolution rather than ADS scenarios, it likewise separates model-guided iteration from execution-based acceptance. Practitioners identify requirements for training data~\cite{yu2024makes}, while ChatGPT can misjudge its code outputs~\cite{yu2024fight}. Code benchmarks~\cite{chen2021evaluating} and repository-level issue resolution~\cite{jimenez2024swebench} therefore require behavioral evaluation. CodaMOSA combines model proposals with search~\cite{lemieux2023codamosa}; TitanFuzz~\cite{deng2023large} and Fuzz4All~\cite{xia2024fuzz4all} generate inputs for execution; ChatUniTest adds validation and repair~\cite{chen2024chatunitest}. ChatScene~\cite{zhang2024chatscene} and LLM4Drive~\cite{yang2023llm4drive} generate driving scenarios, while multimodal approaches are surveyed separately~\cite{cui2024survey}. Generated artifacts still need syntax, execution, and semantic checks. Cyber-physical scenarios additionally impose map, mission, actor-placement, and simulator constraints. \tool adopts this generation--validation separation, using execution evidence to determine regression-family membership.

\section{\tool}
\label{sec:framework}

\subsection{Framework Overview}

\tool is a software-engineering framework for post-fuzzing test amplification. Given a DriveFuzz-discovered failure case, it first organizes the original failure into structured evidence, then uses LLM-assisted causal explanation to derive a root-cause pattern, and finally generates and validates candidate scenarios through rule-based concretization and real DriveFuzz, Autoware, and CARLA execution. Figure~\ref{fig:bug2scenario-overview} illustrates this workflow.

For the running example in Fig.~\ref{fig:motivating-example}, this chain starts from the observed actor--route conflict, retains that interaction in the root-cause pattern and high-level variants, converts each variant into candidate JSON, and accepts it only when execution and manual evidence support the same collision mechanism.

The workflow contains three interacting phases. \textbf{The first phase is semantic understanding.} \tool represents the original seed failure case through scenario data, an oracle result, an execution trace, and root-cause evidence. The failure abstraction module converts these inputs into a scenario summary, critical-window record, oracle summary, and trace evidence. The failure analyzer then uses LLM-assisted causal explanation to identify the failure mechanism and expose reasoning uncertainty. Its output is the root-cause pattern \(P\), which records the mechanism and preservation constraints for later generation.

\textbf{The second phase is grounded generation.} Guided by \(P\), the scenario amplifier produces root-cause-guided, high-level scenario variants that preserve key failure conditions while changing controllable actors, missions, puddles, or weather. These variants are not simulator inputs. The rule-based concretizer performs schema alignment and actor, mission, and weather instantiation to emit a DriveFuzz-style candidate scenario JSON.

\textbf{The third phase is execution-based validation.} The validation suite first applies pre-execution validation, including schema, actor-placement, and mission-feasibility checks. Candidates that pass are executed with DriveFuzz, Autoware, and CARLA, producing oracle results, traces, logs, and rosbags. Post-execution classification assigns each candidate to an invalid, no-failure, different-failure, or same-root-cause category. Manual verification converts the available evidence into validation results: PASS for a same-root-cause failure or FAIL for an invalid, no-failure, or different-failure outcome.

The framework is iterative rather than one-shot. The failure memory \& control module stores historical failure records, constructs prompts, applies rule-based filtering, and limits cost through seed-level process control. Feedback path R1 rejects or revises candidates that fail validation. R2 updates the causal explanation when execution contradicts it, and R3 updates generation constraints after no-failure or different-failure outcomes. R4 stores each validated pattern, while confirmed same-root-cause candidates form the amplification outputs.

\subsection{Problem Formulation}

A DriveFuzz-style scenario configuration can be represented as \(T=\langle Map, Mission, Actors, Puddles, Weather\rangle\). \tool focuses on the post-fuzzing setting, where the input is not an arbitrary seed scenario but a failure case that has already triggered an oracle violation. We denote such a seed failure case as
\begin{equation}
F=\langle T_{\mathrm{fail}}, O, Tr, E\rangle,
\end{equation}
where \(T_{\mathrm{fail}}\) is the original failure-triggering scenario, \(O\) is the observed oracle result---\textbf{collision (C)}, \textbf{red-light violation (R)}, \textbf{lane invasion (L)}, or \textbf{immobility/stuck (S)}---\(Tr\) is the execution trace containing vehicle states and runtime observations, and \(E\) is the available root-cause evidence collected from manual inspection, logs, topic summaries, or LLM-assisted explanation.

The goal of \tool is to generate a set of executable candidate scenarios
\begin{equation}
\mathcal{T}'=\{T'_1,T'_2,\ldots,T'_n\},
\end{equation}
where each candidate scenario is also a DriveFuzz-style configuration
\begin{equation}
T'_i=\langle Map'_i, Mission'_i, Actors'_i, Puddles'_i, Weather'_i\rangle.
\end{equation}
Unlike diversity-oriented scenario generation, our objective is not to maximize arbitrary variation. Instead, each \(T'_i\) should vary the scenario conditions while preserving the failure mechanism captured by the root-cause pattern \(P\) of \(F\). The generated scenario is therefore a candidate amplified scenario until it is validated through execution.

After a generated candidate is concretized and executed, \tool classifies the outcome into four categories. An \textbf{invalid scenario} fails pre-execution checks or cannot be executed reliably in the simulator. A \textbf{no-failure scenario} executes successfully but does not trigger a relevant oracle violation. A \textbf{different-failure scenario} triggers a failure, but the failure symptom or trace evidence does not match the seed root-cause pattern. A \textbf{same-root-cause failure} triggers a compatible oracle result and produces trace evidence consistent with the original root-cause pattern. Only the last category is considered a validated member of the amplified bug family.

\subsection{Failure Abstraction and Causal Explanation Labeling}

The failure abstraction module first converts a discovered failure into information that can support causal explanation. Given a seed failure case \(F\), it constructs an input consisting of the scenario data, fault label, oracle result, and selected execution evidence. The scenario summary covers the map, mission, actors, puddles, and weather parameters, and the oracle summary records the observed outcome. When available, the trace evidence contains critical-window observations, actor states, ego-vehicle states, and relevant log or rosbag information.

The failure analyzer asks the LLM to produce a structured causal explanation rather than a final scenario. A typical explanation includes a failure summary, key entities, candidate causal interactions, reasoning uncertainty, and conditions that should be preserved during amplification. For example, in a collision case, the explanation may identify that the ego vehicle approaches a conflict region, an actor is placed near or along the ego route, the ADS does not avoid or slow down sufficiently, and the execution eventually results in collision. The LLM output is treated as a candidate explanation and not as a ground-truth oracle. We therefore design the prompt to require uncertainty statements and evidence-based reasoning, and we manually inspect a small set of examples before applying the prompt to a larger failure pool through an API.

This design separates semantic interpretation from execution validation. The LLM is useful for summarizing complex scenario information and proposing causal hypotheses, but it does not decide whether a generated scenario is correct. Correctness is checked later through rule-based validation and simulator execution. This separation is important because LLM explanations may be incomplete or may overfit to the fault label if trace evidence and root-cause constraints are not explicitly provided.

\subsection{Root-Cause Pattern Extraction}

After obtaining a causal explanation, \tool abstracts it into a root-cause pattern \(P\). This pattern is more specific than an oracle label. The oracle label describes the failure symptom, while \(P\) describes the mechanism that should be preserved during amplification. For example, the label ``collision'' only states that a collision occurs, whereas \(P\) may capture that an actor partially obstructs the ego route, the ego vehicle enters the conflict region, and the planner or controller fails to avoid or slow down before impact.

In our framework, a root-cause pattern records four types of information. First, it identifies the relevant scenario entities, such as the ego vehicle, actor, puddle, traffic signal, road segment, or intersection. Second, it describes the interaction among these entities, such as route obstruction, unsafe approach, insufficient stopping distance, or failure to re-plan. Third, it specifies the failure symptom that should remain compatible with the seed failure, such as collision, red-light violation, immobility, or a combination of oracle labels. Fourth, it records preservation constraints that guide generation, such as keeping an actor near the ego route, keeping the mission through a conflict region, or retaining a low-friction area near the critical window.

The pattern is not intended to replace detailed manual debugging of Autoware modules. Instead, it acts as an intermediate representation for test amplification. It is sufficiently structured to constrain scenario generation, but lightweight enough to be obtained from scenario data, oracle results, traces, and LLM-assisted explanation. This representation allows \tool to generate scenarios that vary surface-level parameters while maintaining the mechanism that plausibly caused the original failure.

\subsection{Root-Cause-Guided Scenario Amplification}

Given a seed failure and its root-cause pattern, the scenario amplifier generates candidate scenario variants. Unlike pure LLM generation, which may directly propose a new scenario from a short fault label, \tool conditions generation on the root-cause pattern and its preservation constraints. The generation step therefore focuses on controlled variation: it changes scenario factors that can test the boundary of the bug while keeping the causal structure of the failure.

The generated variants are high-level templates rather than final simulator coordinates. A collision pattern can vary actor placement, speed, route geometry, puddles, or weather while retaining the conflict between the ego route and the obstructing actor. For a red-light failure, generation may alter the approach route, controlled intersection, weather, or initial distance to the stop line without removing the relevant signal-handling condition. Stuck patterns instead vary mission geometry, blocking actors, or road conditions while keeping the conditions under which the ADS fails to progress.

This design has two advantages. First, it reduces the chance that the LLM generates arbitrary diverse scenarios unrelated to the seed bug. Second, it provides a natural way to study triggering boundaries. If a generated variant still produces a same-root-cause failure, it becomes part of the bug family. If it no longer triggers a failure, it provides evidence about the boundary of the failure mechanism. Thus, amplification is not only a generation task but also a way to characterize when a known ADS bug persists or disappears.

\subsection{Rule-based Concretization and Validation}

The high-level variants generated by the scenario amplifier must be converted into executable DriveFuzz-style scenario configurations. The rule-based concretizer aligns each generated template with the expected schema, instantiates map and mission fields, places actors and puddles, assigns weather parameters, and produces a candidate scenario JSON \(T'\). This rule-based step is necessary because LLM-generated descriptions alone may contain invalid coordinates, unsupported fields, physically implausible placements, or incomplete execution parameters.

Before executing a candidate, the validation suite applies pre-execution validation. Its schema check, actor-placement check, and mission-feasibility check also test physical plausibility and preservation of key root-cause conditions. For instance, an actor should not be placed outside valid road regions or overlap unrealistically with the ego vehicle at initialization. A mission should remain executable and should still pass through the region relevant to the seed failure. For red-light-related cases, the route should still involve a relevant traffic-light-controlled area. These deterministic checks precede simulator execution, while seed-level stopping limits later API calls after strict recovery.

Candidates that pass pre-execution validation are run in the DriveFuzz, Autoware, and CARLA environment, which produces oracle results, traces, logs, and rosbags. During post-execution classification, an unreliable execution or violated static assumption is marked invalid. A reliable run without a relevant oracle is a no-failure case. A failure inconsistent with the seed pattern is a different-failure case. Only a compatible oracle supported by trace evidence is classified as a same-root-cause failure.

The feedback loops use manual verification and automatic evidence to produce same-seed validation results. Manual verification refines a same-root-cause classification with the conservative label \textbf{strictly preserved}, \textbf{partial or boundary}, or \textbf{not preserved}, together with the observed failure type and notes. The failure memory \& control module retains these records as historical failure storage without replacing the authoritative original seed or its Phase~1 pattern. Rule-based filtering selects relevant records, prompt construction combines them with the original target, and the LLM revises its explanation and proposes new mutations. Process control stops a seed once at least one candidate is strictly preserved to reduce later API and simulator cost. Partial or boundary cases remain feedback evidence but do not satisfy the stopping criterion.

\section{Experiments and Evaluation}

\subsection{Research Questions and Evaluation Overview}

We evaluate \tool as a post-fuzzing scenario amplification framework for Autoware failures discovered by DriveFuzz. The evaluation focuses on execution-grounded amplification rather than textual explanation quality alone. A generated candidate is useful only if it can be represented as a DriveFuzz-style scenario, executed in the DriveFuzz, Autoware, and CARLA environment, and classified with respect to the seed failure's root-cause pattern. The evaluation does not require the generated failures to be uniformly distributed across oracle types. Instead, it uses a balanced generation budget over selected seed failures and model backbones, and treats the resulting failure distribution as an empirical outcome.

\noindent\textbf{RQ1---Backbone effects.} Under a shared generation protocol, how do the six LLM backbones differ in candidate validity and in the execution evidence available for later assessment?

\noindent\textbf{RQ2---Root-cause preservation.} Among executable or oracle-triggering candidates, which ones retain the seed mechanism rather than becoming invalid, no-failure, or different-failure scenarios?

\noindent\textbf{RQ3---Matched Pure-LLM comparison.} A matched control tests whether root-cause guidance changes executable and preservation-consistent outcomes under the same generation and execution budget.

\noindent\textbf{RQ4---Exploratory feedback repair.} For seeds without a strict first-pass success, we follow recovery for at most five rounds under seed-level stopping.

Candidate assessment proceeds through four validation stages. First, \emph{JSON validity} checks whether generated candidate files are syntactically parseable. Second, \emph{execution artifact availability} records whether DriveFuzz produces rosbags, score files, camera videos, or only minimal artifacts. Third, \emph{oracle triggering} identifies an execution-level failure symptom. The final stage determines whether the observed behavior is consistent with the seed root-cause pattern. This ordering matters because syntactic validity does not guarantee simulator execution, and an oracle-triggered execution is not necessarily a same-root-cause failure.

\subsection{DriveFuzz Baseline and Failure Pool}

We use DriveFuzz as the failure-discovery baseline. Among 1,000 executions, 169 trigger at least one failure oracle. Manual audit found 13 of the 169 cases whose oracle-label combinations differed from their DriveFuzz records and annotated 19 as uncertain, motivating our separation of oracle symptoms from mechanism assessment. The pool is also overlapping and imbalanced: it contains 23 collision, 10 stuck, 76 lane-invasion, and 150 red-light labels.

We stratify the overlapping pool, oversampling rare collision, stuck, and mixed-oracle groups while sampling the dominant red-light and lane-plus-red-light groups across the pool. The resulting 30 seeds involve 12 collision, 18 red-light, 18 lane-invasion, and 6 stuck labels; these groups overlap. We allocate the same budget to every seed and model. Each of six LLM backbones generates two candidates per seed, giving 360 \tool candidates. The budget is balanced, but the execution and preservation outcomes are empirical rather than forced to be uniform.

\subsection{Experimental Setup}

The Phase~3 execution uses 30 selected seeds, six LLM backbones, and two candidates per seed-model pair. The backbones are DeepSeek R1, Doubao Seed~2.0 Lite, Claude Sonnet~4.5, Gemini~2.5 Pro, GPT-4o, and GPT-5.1. All candidates are exported as DriveFuzz-style JSON files and executed in the DriveFuzz, Autoware, and CARLA environment. Each candidate has its own output directory. We inspect the generated JSON files, execution directories, rosbag artifacts, score files, camera videos, and error-event JSON files. Code-assisted analysis supplies execution and oracle evidence, while manual review determines root-cause preservation.

For RQ3, we additionally execute a matched Pure-LLM baseline. The baseline uses the same 30 seed failures and six model backbones, but generates one candidate per seed-model pair. The prompt contains only the seed scenario summary and DriveFuzz symptom label. It does not include the root-cause pattern, critical-window evidence, or preservation constraints. This produces 180 Pure-LLM candidates. To keep the execution budget matched, we compare them with the predefined \texttt{cand\_001} subset of \tool, which also contains 180 candidates.

\subsection{Manual Review Protocol}

Root-cause preservation requires evidence beyond oracle labels. We therefore use code-assisted evidence extraction followed by manual assessment of the candidate video and available traces against the seed root-cause pattern. Two computer science Ph.D. students participated in evidence processing and assessment under a shared three-level protocol: \emph{strictly preserved}, \emph{partial or boundary}, and \emph{not preserved}. The conservative primary measure counts a candidate as strict only when the oracle and available trace or video evidence unambiguously support the same seed mechanism. The broader inclusive measure is the set union of strict and partial-or-boundary cases and is reported separately from strict success. Strict preservation requires sufficient visual or trace evidence when artifacts are missing or incomplete.

The matched manual evaluation covers all 180 \tool \texttt{cand\_001} candidates and all 180 Pure-LLM candidates. The \tool labels contain 65 strict, 34 boundary, and 81 not-preserved cases. The Pure-LLM labels contain 53 strict, 27 boundary, and 100 not-preserved cases.

\subsection{RQ1: LLM Backbone Comparison}

\begin{table}[t]
\centering
\caption{Matched execution-level comparison (180 candidates per method). The review pool is a triage metric, not preservation ground truth.}
\label{tab:pure-llm-comparison}
\begin{tabular}{lrr}
\toprule
Metric & \tool & Pure-LLM \\
\midrule
Generated / JSON-valid & 180 & 180 \\
Camera artifacts & 177 & 173 \\
Rosbag artifacts & 101 & 94 \\
Oracle triggered & 94 & 87 \\
Root-cause review pool & 84 & 72 \\
Different failure & 10 & 15 \\
No oracle / incomplete & 78 & 87 \\
Invalid / crash & 8 & 6 \\
\bottomrule
\end{tabular}
\end{table}

All six backbones generate 60 syntactically valid DriveFuzz-style candidates, and every candidate has an execution directory. Across the 360 candidates, 197 produce rosbag and camera artifacts, 126 produce score and camera artifacts without a rosbag, 27 produce camera-only artifacts, and 10 produce minimal artifacts. Artifact completeness differs by model: rosbag counts range from 24 to 38, score counts from 16 to 30, and camera counts from 57 to 59 out of 60. These are execution-availability measures, not root-cause-preservation results.

\begin{table*}[!t]
\centering
\caption{Manual RQ3 results under the matched 180-candidate budget. Cells show the \tool rate (\%) with the percentage-point change from Pure-LLM as a subscript. Bold values indicate positive changes over Pure-LLM, not statistical significance. \emph{Incl.} adds partial or boundary cases to strict preservation. C, R, L, and S denote collision, red-light violation, lane invasion, and stuck. Involvement groups overlap. Multiplicity groups partition the seeds.}
\label{tab:rq3-manual-comparison}
\setlength{\tabcolsep}{2.2pt}
\renewcommand{\arraystretch}{1.16}
\resizebox{\textwidth}{!}{%
\begin{tabular}{lrrrrrrrrrrrrrr}
\toprule
\multirow{2}{*}{Seed-oracle group}
& \multicolumn{2}{c}{{\scriptsize DeepSeek R1}}
& \multicolumn{2}{c}{{\scriptsize Doubao Seed 2.0Lite}}
& \multicolumn{2}{c}{{\scriptsize Claude Sonnet 4.5}}
& \multicolumn{2}{c}{{\scriptsize Gemini 2.5Pro}}
& \multicolumn{2}{c}{{\scriptsize GPT-4o}}
& \multicolumn{2}{c}{{\scriptsize GPT-5.1}}
& \multicolumn{2}{c}{Overall} \\
\cmidrule(lr){2-3}\cmidrule(lr){4-5}\cmidrule(lr){6-7}
\cmidrule(lr){8-9}\cmidrule(lr){10-11}\cmidrule(lr){12-13}
\cmidrule(lr){14-15}
& \multicolumn{1}{c}{Strict} & \multicolumn{1}{c}{Incl.}
& \multicolumn{1}{c}{Strict} & \multicolumn{1}{c}{Incl.}
& \multicolumn{1}{c}{Strict} & \multicolumn{1}{c}{Incl.}
& \multicolumn{1}{c}{Strict} & \multicolumn{1}{c}{Incl.}
& \multicolumn{1}{c}{Strict} & \multicolumn{1}{c}{Incl.}
& \multicolumn{1}{c}{Strict} & \multicolumn{1}{c}{Incl.}
& \multicolumn{1}{c}{Strict} & \multicolumn{1}{c}{Incl.} \\
\midrule
\rowcolor{gray!15}\multicolumn{15}{c}{\textit{By oracle involvement (overlapping groups)}} \\
C involved
& \drop{\phantom{0}8.3}{-16.7} & \drop{16.7}{-\phantom{0}8.3}
& \same{16.7} & \drop{16.7}{-\phantom{0}8.3}
& \gain{16.7}{\phantom{0}8.3} & \gain{41.7}{25.0}
& \drop{\phantom{0}0.0}{-\phantom{0}8.3} & \same{\phantom{0}8.3}
& \gain{25.0}{\phantom{0}8.3} & \gain{41.7}{16.7}
& \drop{\phantom{0}0.0}{-16.7} & \same{25.0}
& \drop{11.1}{-\phantom{0}4.2} & \gain{25.0}{\phantom{0}4.2} \\
R involved
& \gain{38.9}{16.7} & \gain{50.0}{11.1}
& \gain{38.9}{11.1} & \gain{50.0}{16.7}
& \gain{50.0}{\phantom{0}5.6} & \gain{72.2}{11.1}
& \drop{27.8}{-\phantom{0}5.6} & \gain{55.6}{11.1}
& \gain{50.0}{16.7} & \gain{83.3}{33.3}
& \gain{33.3}{16.7} & \gain{66.7}{33.3}
& \gain{39.8}{10.2} & \gain{63.0}{19.4} \\
L involved
& \same{16.7} & \gain{27.8}{11.1}
& \same{11.1} & \gain{16.7}{\phantom{0}5.6}
& \drop{22.2}{-11.1} & \same{38.9}
& \gain{16.7}{\phantom{0}5.6} & \gain{27.8}{11.1}
& \gain{22.2}{11.1} & \gain{50.0}{22.2}
& \gain{33.3}{16.7} & \gain{66.7}{38.9}
& \gain{20.4}{\phantom{0}3.7} & \gain{38.0}{14.8} \\
S involved
& \gain{50.0}{16.7} & \gain{50.0}{16.7}
& \same{33.3} & \same{33.3}
& \same{33.3} & \gain{66.7}{16.7}
& \same{33.3} & \gain{66.7}{33.3}
& \gain{33.3}{16.7} & \gain{50.0}{16.7}
& \gain{50.0}{50.0} & \gain{83.3}{33.3}
& \gain{38.9}{13.9} & \gain{58.3}{19.4} \\
\midrule
\rowcolor{gray!15}\multicolumn{15}{c}{\textit{By oracle multiplicity (disjoint groups)}} \\
Single oracle
& \drop{58.3}{-\phantom{0}8.3} & \drop{66.7}{-25.0}
& \same{75.0} & \drop{83.3}{-\phantom{0}8.3}
& \gain{75.0}{41.7} & \gain{91.7}{25.0}
& \drop{50.0}{-\phantom{0}8.3} & \drop{75.0}{-\phantom{0}8.3}
& \gain{75.0}{16.7} & \same{91.7}
& \gain{41.7}{25.0} & \same{75.0}
& \gain{62.5}{11.1} & \drop{80.6}{-\phantom{0}2.8} \\
Two oracles
& \same{16.7} & \gain{33.3}{16.7}
& \gain{16.7}{\phantom{0}8.3} & \gain{25.0}{16.7}
& \drop{33.3}{-\phantom{0}8.3} & \same{50.0}
& \same{16.7} & \gain{33.3}{16.7}
& \gain{25.0}{\phantom{0}8.3} & \gain{50.0}{16.7}
& \gain{41.7}{16.7} & \gain{83.3}{50.0}
& \gain{25.0}{\phantom{0}4.2} & \gain{45.8}{19.4} \\
Three oracles
& \gain{16.7}{16.7} & \gain{16.7}{16.7}
& \same{\phantom{0}0.0} & \same{\phantom{0}0.0}
& \drop{\phantom{0}0.0}{-16.7} & \gain{33.3}{16.7}
& \same{\phantom{0}0.0} & \gain{16.7}{16.7}
& \gain{16.7}{16.7} & \gain{50.0}{50.0}
& \same{\phantom{0}0.0} & \gain{16.7}{16.7}
& \gain{\phantom{0}5.6}{\phantom{0}2.8} & \gain{22.2}{19.4} \\
\midrule
\textbf{Overall}
& \same{33.3} & \same{43.3}
& \gain{36.7}{\phantom{0}3.3} & \gain{43.3}{\phantom{0}3.3}
& \gain{43.3}{10.0} & \gain{63.3}{13.3}
& \drop{26.7}{-\phantom{0}3.3} & \gain{46.7}{\phantom{0}6.7}
& \gain{43.3}{13.3} & \gain{66.7}{16.7}
& \gain{33.3}{16.7} & \gain{66.7}{23.3}
& \gain{36.1}{6.7} & \gain{55.0}{10.6} \\
\bottomrule
\end{tabular}%
}
\end{table*}

Taken together, all six backbones generate JSON-valid candidates, but their execution artifacts are not equally complete. Rosbags, score files, and camera streams capture different portions of an execution and should not be treated as interchangeable success indicators. The nearly uniform camera counts would hide the much wider variation in rosbag and score availability, while a score file without a rosbag cannot support the same temporal inspection. Backbone evaluation must therefore include the evidence needed by downstream preservation assessment rather than stop at syntactic validity or aggregate all artifact types into a single completeness flag.

RQ1 therefore reveals differences after generation rather than at the JSON boundary. Every backbone satisfies the syntactic requirement, yet the evidence available for later diagnosis varies across traces, scores, and videos. A model that produces more camera files is not necessarily the one that leaves the strongest temporal record or triggers the most relevant oracle events. We consequently avoid declaring a single best backbone from artifact counts alone and carry the available evidence into preservation assessment.

\subsection{RQ2: Root-Cause Preservation}

RQ2 evaluates the central claim of \tool: generated scenarios should preserve the seed failure mechanism. A candidate is not counted as successful merely because it triggers any DriveFuzz oracle. It must trigger behavior compatible with the seed oracle and satisfy the root-cause preservation constraints extracted from the seed failure.

Execution outcomes demonstrate the gap between syntax, oracle symptoms, and preservation. Of the 360 \tool candidates, 178 trigger at least one DriveFuzz oracle, 143 produce partial artifacts without an observed oracle, 19 reach the goal without a failure, 10 fail during actor spawning, and 10 terminate with HANG behavior. The available event files contain 36 crash and 152 red-light flags, but these overlapping symptoms do not establish the seed mechanism.

Code-assisted analysis compares each candidate with its seed using oracle compatibility, mission and map preservation, actor-route relations, runtime distance, and extra symptoms. It is used for evidence extraction and triage, not as preservation ground truth. In the fully reviewed 180-candidate matched \tool subset, 65 candidates are strictly preserved, 34 are partial or boundary, and 81 are not preserved. This corresponds to 36.1\% strict preservation and 55.0\% inclusive preservation relevance.

Strict preservation is substantially narrower than oracle triggering: only 65 of the 180 manually reviewed matched candidates are strictly preserved. This gap is expected because an oracle records a symptom, whereas preservation asks whether the interactions that produced that symptom remain compatible with the seed mechanism. The partial or boundary label captures cases in which some defining evidence survives but the full causal pattern cannot be confirmed. Counting those cases changes the scientific claim from reproduced mechanism to preservation-relevant evidence. The inclusive measure is therefore reported separately and never substituted for the conservative strict result.

The three-way manual distribution localizes where evidence is lost. The 34 boundary cases are neither equivalent to the 65 strict reproductions nor appropriately merged with the 81 not-preserved outcomes: they retain some seed-defining relation but lack the complete evidence required by the conservative rule. Keeping this middle category prevents uncertainty from being converted into success or failure. Accordingly, the strict measure records confirmed reproduction, whereas the inclusive measure asks whether generation retained preservation-relevant evidence that may still support diagnosis or targeted repair.

\begin{figure*}[!t]
\centering
\begin{minipage}[t]{0.485\textwidth}
\centering
\includegraphics[width=0.90\linewidth]{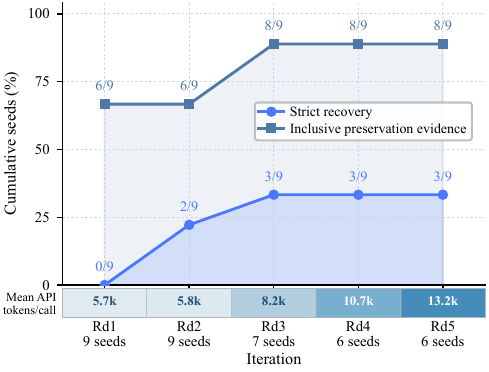}
\captionof{figure}{Final cumulative seed-level RQ4 trajectory with round-level API cost. Plot labels Rd1--Rd5 denote experimental rounds, distinguishing them from feedback paths R1--R4 in Fig.~\ref{fig:bug2scenario-overview}. Both curves use the original nine-seed denominator. Inclusive preservation evidence is the union of strict and partial/boundary outcomes. The bottom strip reports mean total API tokens per call, not round totals.}
\label{fig:rq4-feedback-trajectory}
\end{minipage}\hfill
\begin{minipage}[t]{0.485\textwidth}
\centering
\includegraphics[width=\linewidth]{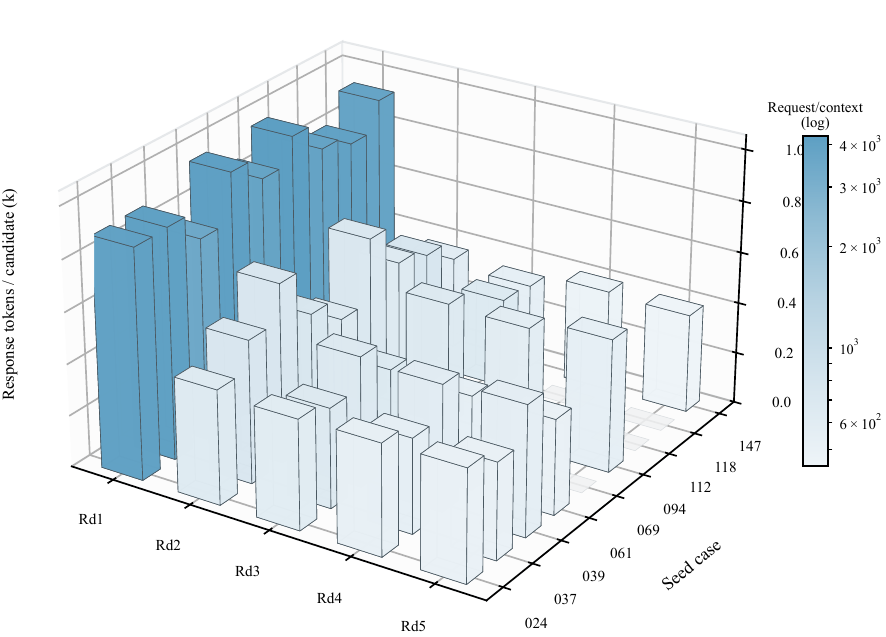}
\captionof{figure}{Seed--round token distribution for the standardized two-candidate audit. Rd1--Rd5 denote experimental rounds. Figure~\ref{fig:rq4-feedback-trajectory} reports round means. Here bar height is response tokens per generated candidate and color is request tokens per actual context record (log scale). Pale tiles denote calls omitted after seed-level stopping.}
\label{fig:rq4-token-normalized}
\end{minipage}
\end{figure*}

\subsection{RQ3: Comparison with Pure-LLM Generation}
RQ3 compares \tool with direct Pure-LLM generation under a matched execution budget. The Pure-LLM baseline receives only the seed scenario summary and fault label, then generates candidate scenarios without the structured root-cause pattern, critical-window evidence, or preservation constraints used by \tool. Both methods use the same seed cases, model backbones, rule-based concretizer, DriveFuzz-style JSON schema, and DriveFuzz, Autoware, and CARLA execution environment.

The RQ3 budget is split evenly between the two methods: 30 seed failures \(\times\) 6 model backbones \(\times\) 1 candidate per seed-model pair, yielding 180 Pure-LLM candidates and 180 matched \tool candidates. The matched \tool subset is the predefined \texttt{cand\_001} candidate from each seed-model pair. Table~\ref{tab:pure-llm-comparison} reports the execution-level comparison, while Table~\ref{tab:rq3-manual-comparison} reports the final manual same-root-cause comparison.

The matched comparison shows that \tool produces slightly more oracle-triggered candidates than Pure-LLM generation under the same budget: 94 versus 87. It also produces more candidates that enter the root-cause review pool: 84 versus 72. Pure-LLM produces more different-failure outcomes, 15 versus 10, and more no-oracle or incomplete outcomes, 87 versus 78. The invalid/crash count is similar, with 8 for the matched \tool subset and 6 for Pure-LLM. These execution-level counts are triage signals rather than preservation ground truth. The primary statistical comparison uses the manual strict and inclusive labels below.

The final manual table reports heterogeneous strict-preservation effects across backbones: the change is zero for DeepSeek, negative for Gemini, and positive for the other four models.

At the execution and oracle-match level, \tool shows a positive trend over Pure-LLM generation under the matched 180-candidate budget. The manual labels narrow this claim. Under the conservative strict criterion, \tool confirms 65 same-root-cause candidates, while Pure-LLM confirms 53. Under the inclusive preservation-relevant criterion, which additionally counts partial or boundary-preserving cases, \tool yields 99 candidates and Pure-LLM yields 80. A two-sided exact McNemar test over the same 180 case-model pairs gives \(p=0.104\) for strict same-root-cause success and \(p=0.011\) for inclusive preservation relevance. Thus, the strict result is positive but not statistically significant at \(p<0.05\), while the inclusive preservation-relevance result shows a significant improvement.

Table~\ref{tab:rq3-manual-comparison} uses 180 candidates per method. Each colored subscript is a percentage-point change from Pure-LLM, not a fraction. Strict requires complete confirmation of the seed mechanism. Inclusive additionally retains partial or boundary evidence. The paired tests therefore compare the same case--model units under both criteria.

Table~\ref{tab:rq3-manual-comparison} is based on 30 seed cases. Applying the same six backbones to these cases yields 180 candidates for each method. In the overlapping view, C, R, L, and S occur in 12, 18, 18, and 6 cases, respectively. By oracle multiplicity, the evaluation set contains 12 single-oracle cases, 12 two-oracle cases, and 6 three-oracle cases. These counts refer to seed cases rather than model-expanded candidate rows.

\subsection{RQ4: Exploratory Feedback-loop Repair}

RQ4 explores the feedback loops in Fig.~\ref{fig:bug2scenario-overview} on the nine seeds for which the first-pass, six-backbone evaluation produced no strictly preserved candidate. We fix the feedback model to Claude Sonnet~4.5 and apply a seed-level stopping rule: a seed leaves the study after its first strictly preserved candidate, whereas partial or boundary cases remain in the feedback history but do not count as recovery. Accordingly, RQ4 is interpreted as an exploratory recovery study for difficult seeds.

Round~2 provides the model with the authoritative original seed and Phase~1 root-cause pattern together with the six same-seed Round~1 candidate, execution, and manual-review records. It generates two candidates for each of the nine seeds. All 18 candidates pass pre-validation and produce execution records. Sixteen have complete front/rear video pairs, while two crash or produce only minimal artifacts. Assessment using the available evidence identifies two strictly preserved candidates, recovering \texttt{case\_069} and \texttt{case\_112}.

Round~3 processes only the remaining seven seeds. Each prompt retains the original target and adds the two Round~2 outcomes to the six Round~1 records. The model generates six candidates per remaining seed, yielding 42 candidates. All are generated, pre-validated, and executed. Of these, 39 have complete video pairs, and all 42 are adjudicated from the available evidence. One candidate strictly preserves the root cause of \texttt{case\_118}. Two additional candidates are labeled partial or boundary and therefore do not satisfy the stopping rule.

Round~4 processes the remaining six seeds and generates 36 candidates. Manual assessment finds no strictly preserved candidate and two partial or boundary candidates, so all six seeds continue. Round~5 generates 36 final candidates. The completed workbook labels the 30 video-reviewed candidates as 0 strict, 3 partial or boundary, and 27 not preserved. Three additional executions have crash or minimal artifacts, and three have no execution output. None is counted as preservation evidence. Thus, Round~5 adds no strict recovery and no new seed to the cumulative evidence set.

Figure~\ref{fig:rq4-feedback-trajectory} combines outcome progression with mean total tokens per API call. Its values are not round totals. Round~1 pools six backbones, whereas Rounds~2--5 use Claude Sonnet~4.5, so the strip is descriptive rather than a cross-model efficiency comparison. The actual active-seed schedule requires $9+7+6+6=28$ feedback calls, compared with $9\times4=36$ calls if all nine seeds continued through four rounds; seed-level stopping therefore avoids eight calls (22.2\%). This result quantifies API-call savings; a complete end-to-end cost analysis additionally requires human-review effort, wall-clock time, and monetary cost.

For the standardized candidate-level audit in Fig.~\ref{fig:rq4-token-normalized}, we retain exactly two outcomes per active seed-round under a rule fixed independently of token usage: strict outcomes first, then partial or boundary outcomes, and finally candidate order. This cap preserves every strict and partial/boundary seed-level finding. Excluded rows remain in the artifact. Token accounting nevertheless follows the actual API calls. The original failed seed and its Phase~1 representation count as one context object and each historical feedback record as one more, giving actual per-call context sizes of 1, 7, 9, 15, and 21 for Rounds~1--5. We normalize request tokens by actual context size and response tokens by the number of candidates actually generated. Thus, post-hoc retention does not erase the cost of excluded outputs.

The composition of the difficult pool offers an exploratory explanation for the plateau. All nine seeds combine multiple oracle types: five contain two oracles and four contain three, whereas the full 30-seed set contains 12 single-, 12 two-, and six three-oracle seeds. Three-oracle seeds are therefore overrepresented in the difficult pool (44.4\% versus 20.0\%), and no single-oracle seed enters it. Round~2 recovers the two C+L seeds \texttt{case\_069} and \texttt{case\_112}. Round~3 recovers \texttt{case\_118} (C+L+R). The six unrecovered seeds are evenly split between two- and three-oracle groups. Their evidence is also harder to resolve: for example, \texttt{case\_037} and \texttt{case\_061} carry C+L+R labels but contain conflicts between the dataset and execution oracle records, while \texttt{case\_094} carries C+L+S although its execution record confirms only collision. These cases also exhibit command--status mismatches that prevent confident module-level attribution. We therefore treat the observed association between compound evidence and recovery difficulty as exploratory.

Across the feedback study, execution and manual-review feedback recovers three of the nine initially unsuccessful seeds, all by Round~3. Round~2 recovers 2/9 seeds, and Round~3 recovers one of the seven remaining seeds, giving a final cumulative recovery of 3/9 (33.3\%). Rounds~4 and 5 produce additional partial or boundary candidates but no strict recovery. The final cumulative preservation-evidence count remains 8/9. The plateau indicates that longer histories alone do not resolve contradictory or incomplete evidence. A more targeted repair stage should first distinguish placement or mission infeasibility from genuine mechanism loss, then revise the corresponding constraint rather than repeatedly enlarging the prompt. Thus, same-seed feedback repaired three initially unsuccessful seeds under this exploratory setting.

\subsection{Cross-RQ Model and Evidence Effects}

Artifact availability and preservation vary across backbones without forming a single model ranking. Camera evidence is nearly complete for every model, whereas rosbag, score-file, and oracle availability differ more clearly. Strict-preservation effects vary by backbone, while inclusive gains are more consistent, particularly for compound-oracle seeds that often retain only partial or boundary evidence. Because subgroup sizes are small and oracle-involvement groups overlap, we interpret these patterns descriptively and report artifact completeness and preservation separately rather than collapse them into a composite score.

\subsection{Error Analysis}

The 360-candidate \tool pool exposes three recurring modes: 10 actor-spawn failures, 10 HANG terminations, and 143 partial artifacts without an oracle. The matched comparison likewise contains no-oracle, incomplete, invalid, and crash outcomes for both methods. None supplies same-root-cause evidence. These modes justify validation beyond JSON syntax, including actor placement, mission feasibility, trace availability, oracle compatibility, and manual preservation assessment. Feedback should therefore target the observed execution failure rather than generic diversity.

\section{Discussion}

\subsection{From Failure Discovery to Regression-Test Amplification}

\tool complements DriveFuzz by beginning where fuzz-driven failure discovery ends. It extracts a root-cause pattern from a discovered failure, generates controlled regression variants, and separates symptom recurrence from mechanism preservation using execution evidence.

This division of software-testing responsibilities is important. DriveFuzz explores the scenario space and exposes failures. \tool turns selected failures into auditable regression targets. A root-cause pattern remains stable across variations, while the validation hierarchy prevents executable but unrelated violations from entering the same bug family. The resulting family supports debugging, regression testing, and test maintenance by showing which controlled changes preserve, weaken, or eliminate a mechanism. It supplements rather than replaces the original fuzzing campaign, and the resulting family remains traceable to its seed failure.

After an ADS update, engineers can rerun the generated family to determine whether the original mechanism persists, weakens, or disappears across controlled scenario variations.

\subsection{Validator-Centered Use of LLMs}

LLMs summarize evidence, propose causal hypotheses, and generate high-level test variants. Rule-based checks and DriveFuzz--Autoware--CARLA evidence decide executability and mechanism preservation. The LLM is therefore a test-hypothesis generator rather than an execution oracle. Seed evidence constrains its explanation, and deterministic schema, placement, and mission logic translate each proposal before simulation. This LLM-assisted but validator-centered design limits fluent unsupported details.

\subsection{Evidence Operationalization}

An available oracle timestamp anchors a bounded critical window. We summarize evidence within or immediately before that window, including ego motion and control, detected or predicted actors, route state, and the oracle event. Large topics are represented by bounded summaries rather than raw arrays, and dataset--execution conflicts remain explicit uncertainty. For \texttt{case\_098}, the collision timestamp anchors the window; the record includes pre-collision speed and braking status, nearby-object evidence, and route waypoints. These fields support the actor--route interpretation in Fig.~\ref{fig:motivating-example} without treating topic availability alone as causal proof.

Pattern construction then uses a fixed evidence-facing template: entities, interaction chain, symptom, and preservation constraints. A proposed element must be traceable to scenario or execution evidence; unsupported elements remain in the uncertainty field rather than becoming hard constraints. The normalized pattern is the contract between semantic interpretation and generation, while the original seed evidence remains authoritative. Together with the concretizer interface in Section~III-F, this makes the transformation boundary auditable.

This representation exposes the evidence supporting each preservation constraint and provides a stable interface between semantic analysis and executable scenario construction.

The concretizer interface follows the same five-part representation. It maps mission poses, actor type, navigation mode, speed and pose, optional actor destinations, puddle location and size, and weather values into named DriveFuzz fields. Missing or unsupported fields are rejected before execution rather than completed by the LLM, exposing the implemented schema boundary for audit.

Each candidate therefore remains traceable to its high-level variant and seed pattern, allowing validation failures to be attributed to a specific transformation stage.

\subsection{Threats to Validity}

\textit{Construct validity.} Regression-family membership depends on seed-derived patterns and manual interpretation of available execution evidence. We retain videos, traces, labels, and matched case--model pairs so that the three-way judgments remain auditable, and we keep strict and inclusive outcomes separate to distinguish boundary evidence from confirmed reproduction. The present study evaluates the concretization pipeline as part of the complete framework; future formal schema-to-JSON mappings and component ablations will quantify its standalone contribution.

\textit{Internal and reliability validity.} Manual assessment under the shared protocol may still reflect judgment variation. Prompt design and backbone choice may influence generated variants. Simulator instability, incomplete videos or traces, and conflicts among dataset and execution-oracle records may also affect classification. Code-assisted triage supports the conservative manual decision rule used for final classification.

\textit{External validity.} The evaluation covers 30 selected DriveFuzz seeds, Autoware and CARLA, selected maps and oracle combinations, and six LLM backbones. The subgroup and difficult-pool observations are interpreted descriptively because groups overlap and sizes differ. RQ4 provides exploratory recovery evidence for nine initially unsuccessful seeds and Claude Sonnet~4.5; broader seed and backbone coverage remains future work. The public repository releases protocols, labels, code, and analysis artifacts to support independent annotation and replication across additional traffic settings and autonomous software stacks.

\section{Conclusion}

This paper presented \tool, which converts a fuzz-discovered ADS failure into a controlled family of regression scenarios by separating root-cause interpretation, deterministic DriveFuzz test construction, and execution-based preservation assessment. Under the matched budget, boundary-inclusive preservation improves by 10.6 percentage points ($p=0.011$), while strict preservation rises by 6.7 points without statistical significance ($p=0.104$). Feedback recovers three of nine initially unsuccessful seeds by Round~3, and seed-level stopping avoids eight of 36 possible feedback calls. These results show that evidence-grounded amplification can produce reusable regression assets while keeping mechanism-preservation judgments auditable. Future work will extend feedback evaluation across more seeds and LLM backbones and refine schema mappings, trace-derived patterns, actor placement, and mission-feasibility repair.

\section*{Acknowledgement} The work described in this paper was fully supported by Hong Kong Metropolitan University Research Grant (Project Reference No. RD/2026/1.15).

\newpage
\IEEEtriggeratref{19}
\bibliographystyle{IEEEtran}
\bibliography{references}

\end{document}